\documentclass[aps,preprint]{revtex4}
\usepackage{epsfig}

\begin{document}

\title{Higher Cumulants in the Cluster Expansion in QCD\footnote{
This paper is dedicated to Franz Wegner on the occasion of his 60th birthday
} }

\author{W. Kornelis and H.G. Dosch}

\affiliation{\centering Institut f\"ur Theoretische Physik\\
Universit\"at Heidelberg\\
Philosophenweg 16\\
D-69120 Heidelberg, Germany}

\begin{abstract}
\mdseries In this work an extension of the Gaussian model of the stochastic
vacuum is presented. It consists of including higher cumulants than just the
second one in the cluster expansion in QCD. The influence of nonabelian fourth
cumulants on the potential of a static quark-antiquarkpair is examined and the
formation of flux tubes between a static \( q\bar{q} \)-pair is investigated.
It is found that the fourth cumulants can contribute to chromomagnetic flux
tubes. Furthermore the contribution of fourth cumulants to the total cross section
of soft high energy hadron-hadron scattering is examined and it is found that
the fourth cumulants do not change the general picture obtained in the Gaussian
model.

\end{abstract}

\maketitle

\section{Introduction}

The linked cluster expansion \cite{kampen1}, \cite{kampen2} is a powerful
tool in statistical mechanics to evaluate expectation values. Especially if
the measure for forming the average is not known it allows a systematic expansion.
In particle physics this method is much less widely used. Here the weight, given
by the exponential function of the action is known at least formally and therefore
there are two principle ways to evaluate expectation values: 

The perturbative one where the action is split in a part quadratic in the fields,
and the rest, which is expanded as a power series. The Gaussian integral due
to the quadratic part can be performed and this yields the well known perturbation
series \cite{Dir33}, \cite{Fey48}.

Another way is to discretize the action in a gauge invariant way on the lattice
\cite{wegner1}, \cite{wilson2} and perform the functional integration with
numerical methods. 

In the model of the stochastic vacuum (MSV) \cite{DS88}, \cite{simonov1},
\cite{Dos87} the linked cluster expansion of the long distance behaviour of
QCD is the starting point. It turns out that if the stochastic variables are
taken to be the gluon field strenghts a convergent cluster expansion yields
automatically the salient feature of non-perturbative QCD namely confinement.
This approach has been successfully applied to many processes and it turned
out that the simplest form of the cluster expansion namely the one where all
cumulants higher than the second one are neglected yields a surprisingly good
description of many phenomena ranging from charmonium spectroscopy to high energy
scattering (see the reviews \cite{Dos94}, \cite{Sim96}, \cite{Dos96}, \cite{Nac97}
and the literature quoted there). A process with vanishing higher cumulants
is just a Gaussian one where the functional integrals can be performed explicitly
and all can be reduced to one correlator. 

Though the Gaussian approximation is astonishingly successful, it is nevertheless
worthwhile to investigate which principally new effects can occur if higher
cumulants are taken into account. This is the aim of the present paper. If one
wants to study the gauge invariant field contribution of a static quark-antiquark
pair and high energy scattering of hadrons, one has to evaluate the vacuum expectation
value of a product of traces of Wegner-Wilson loops. In that case the leading
term is the expectation value of the product of four field strength tensors.
A fourth cumulant can contribute directly to that expectation value and for
that reason in this paper we concentrate on the fourth cumulant. 

It is structured in the following way: In the second section we shortly repeat
some basic features of the linked cluster expansion and some important technical
features of the model of the stochastic vacuum (MSV). In section \ref{highercumuls}
we discuss some possible choices of higher cumulants and the way how we calculate
the effects of these higher cumulants. We present the results obtained in this
way and discuss them.

\section{Linked Cluster Expansion and the Model of the Stochastic Vacuum}

The quantity we are mainly interested in, is the Wegner-Wilson loop in QCD \cite{wegner1},
\cite{wilson2}. 

It is defined as the vacuum expectation value of the gauge invariant path ordered
closed line integral over the exponential of the color-potential
\begin{equation}
\label{wwloop}
\left\langle \mathbf{W}[C]\right\rangle =\big <\mathrm{P}e^{-ig\oint _{C}\mathbf{A}_{\mu }(x)\, dx^{\mu }}\big >
\end{equation}
 where \( \mathbf{A}_{\mu }(z)=\sum A_{\mu }^{a}\lambda ^{a}/2 \) is the SU(3)-Lie-Algebra
valued gluon potential and \( \mathrm{P} \) denotes path ordering. 

\begin{figure}
\centering\begin{minipage}{10cm}\centering

{\par\centering \resizebox*{10cm}{!}{\includegraphics{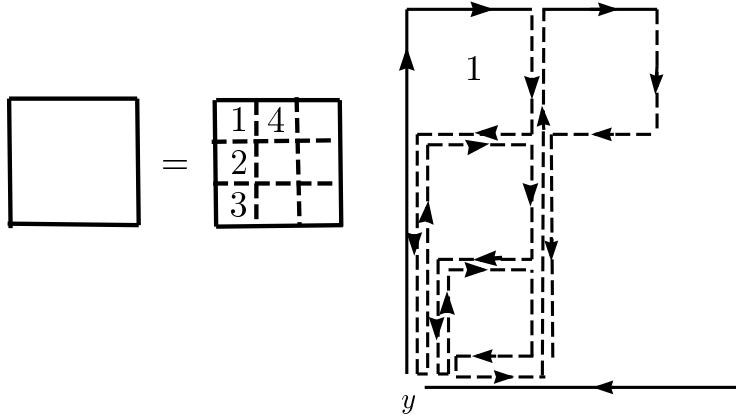}} \par}

\caption[Surface ordering]{\label{stokes}Transition of path ordering to surface
ordering with the non-abelian Stokes theorem.}

\end{minipage}
\end{figure}

Using the non-Abelian Stokes theorem \cite{Are80}, \cite{Bra80}, \cite{Sim88},
we can express \( \mathbf{W}[C] \) by a surface integral over the gluon field
strength
\begin{equation}
\label{wwloopsurface}
\left\langle \mathbf{W}[C]\right\rangle =\big <\mathrm{P}_{S}e^{-ig\frac{1}{2}\int _{S}\mathbf{F}_{\mu \nu }(x)\, d\sigma ^{\mu \nu }(x)}\big >
\end{equation}
where the border of \( S \) is the closed loop \( C \) and \( d\sigma ^{\mu \nu } \)
is the usual oriented surface element of \( S \) at \( x \); we choose in
the sums always \( \mu <\nu  \). The path ordering of equation (\ref{wwloop})
becomes here a surface ordering implied by the non-Abelian Stokes theorem. A
possible surface ordering is given in fig. \ref{stokes}
. If we say in the following \( x_{1}<x_{2}\ldots <x_{n} \) we mean ordered
according to surface ordering. In an Abelian theory (electrodynamics) the field
strength tensor is a matrix of \( C \)-numbers and no path ordering is necessary.
Then (\ref{wwloopsurface}) is just the familiar Stokes theorem in 4 dimensions. 

The linked clusters are now defined by: 
\begin{eqnarray}
\lefteqn {\left\langle \mathbf{W}[C]\right\rangle =\mathrm{P}_{S}\exp \Big [-\sum _{n}\frac{g^{n}}{2^{n}n!}} &  & \nonumber \\
 &  & \int _{S}d\sigma _{\mu _{1}\nu _{1}}(x_{1})\ldots d\sigma _{\mu _{n}\nu _{n}}(x_{n})\, \left< \left< \, \mathbf{F}_{\mu _{1}\nu _{1}}(x_{1})\ldots \mathbf{F}_{\mu _{n}\nu _{n}}(x_{n})\, \right> \right> \Big ]
\end{eqnarray}
Lorentz and gauge invariance imply \( \left\langle \left< \, \mathbf{F}_{\mu \nu }(x)\, \right> \right\rangle =0 \)
and the series starts with \( n=2 \), the ``Gaussian'' cluster. 

By expanding the exponentials and rearranging the terms one can show that the
cumulants \( \left\langle \left\langle .\right\rangle \right\rangle  \) can
be expressed through the expectation values \( \left\langle .\right\rangle  \)
\cite{kampen2}. 
\begin{eqnarray}
\left\langle \left\langle F_{1}F_{2}\right\rangle \right\rangle  & = & \left\langle F_{1}F_{2}\right\rangle \quad \left\langle \left\langle F_{1}F_{2}F_{3}\right\rangle \right\rangle =\left\langle F_{1}F_{2}F_{3}\right\rangle 
\end{eqnarray}
and for \( x_{1}<x_{2}\ldots <x_{n} \)
\begin{eqnarray}
\left\langle \left\langle F_{1}F_{2}F_{3}F_{4}\right\rangle \right\rangle  & = & \left\langle F_{1}F_{2}F_{3}F_{4}\right\rangle -\left\langle F_{1}F_{2}\right\rangle \left\langle F_{3}F_{4}\right\rangle -\left\langle F_{1}F_{3}\right\rangle \left\langle F_{2}F_{4}\right\rangle -\left\langle F_{1}F_{4}\right\rangle \left\langle F_{2}F_{3}\right\rangle \nonumber \\
 & \vdots  & 
\end{eqnarray}
where \( F_{i}=F_{\mu _{i}\nu _{i}}(x_{i}) \). 

For a Gaussian process all cumulants higher than the second one vanish by definition
and we obtain directly for such a process
\begin{eqnarray}
\lefteqn {\left\langle \mathbf{W}[C]\right\rangle =\mathrm{P}_{S}} &  & \nonumber \\
 &  & \exp \Big [-\frac{g^{2}}{2^{2}2!}\int _{S}\int _{S}d\sigma _{\mu _{1}\nu _{1}}(x_{1})d\sigma _{\mu _{2}\nu _{2}}(x_{2})\, \left< \left< \, \mathbf{F}_{\mu _{1}\nu _{1}}(x_{1})\mathbf{F}_{\mu _{2}\nu _{2}}(x_{2})\, \right> \right> \Big ]
\end{eqnarray}
from which the area law of the W-loop can be derived easily \cite{Dos87}, \cite{DS88}
if we assume that the correlator itself falls off with a characteristic correlation
length \( a \). 

In order to evaluate expectation values of two (or more) Wegner-Wilson loops
the simple factorization of the matrices does not suffice. In all applications
studied so far, the following factorization hypothesis has been made for the
color components of the field
\begin{eqnarray}
\left\langle F_{1}^{a_{1}}F_{2}^{a_{2}}F_{3}^{a_{3}}F_{4}^{a_{4}}\right\rangle  & = & \left\langle F_{1}^{a_{1}}F_{2}^{a_{2}}\right\rangle \left\langle F_{3}^{a_{3}}F_{4}^{a_{4}}\right\rangle \nonumber \\
 &  & +\left\langle F_{1}^{a_{1}}F_{3}^{a_{3}}\right\rangle \left\langle F_{2}^{a_{2}}F_{4}^{a_{4}}\right\rangle +\left\langle F_{1}^{a_{1}}F_{4}^{a_{4}}\right\rangle \left\langle F_{2}^{a_{2}}F_{3}^{a_{3}}\right\rangle 
\end{eqnarray}
with
\begin{equation}
\label{nonlocalcorr}
F_{i}^{a_{i}}=F^{a_{i}}_{\mu _{i}\nu _{i}}(x_{i},w)
\end{equation}
\( \mathbf{F}_{\mu _{i}\nu _{i}}(x_{i},w) \) is the color-parallel transported
field:
\begin{equation}
\label{transport}
\mathbf{F}_{\mu _{i}\nu _{i}}(x_{i},w)=\Phi ^{-1}(x_{i},w)\mathbf{F}_{\mu \nu }(x_{i})\Phi (x_{i},w)
\end{equation}
with
\begin{equation}
\Phi (x_{i},w)=\mathrm{P}\exp \left[ -ig\int _{0}^{1}d\lambda \, \mathbf{A}_{\mu }\left( w+\lambda (x_{i}-w)\right) \, (x_{i}^{\mu }-w^{\mu })\right] 
\end{equation}
being the color-parallel transporter from \( x \) to \( w \) in the adjoint
representation. Transformation (\ref{transport}) is necessary in order to give
the non-local correlators a gauge invariant meaning. 

In the applications of the MSV to the calculations of more than one W-loop the
following approximations are made: In the correlator \( \left< \left< \, F^{a_{1}}_{\mu _{1}\nu _{1}}(x_{1},w)F^{a_{2}}_{\mu _{2}\nu _{2}}(x_{2},w)\, \right> \right>  \)
the dependence on the reference point \( w \) is neglected which is certainly
only justifiable for a ``reasonable'' choice of the point \( w \). A more
technical assumption is the following: if the exponentials are expanded as power
series in the correlator \( \left< \left< \, F^{a_{1}}_{\mu _{1}\nu _{1}}(x_{1},w)F^{a_{2}}_{\mu _{2}\nu _{2}}(x_{2},w)\, \right> \right>  \)
only the term leading in \( (a/L) \) is taken into account, where \( L \)
is the linear dimension of the loop and \( a \) the correlation length of the
correlator. For a discussion and possible resolution of this rather ugly assumption
see \cite{RD95}, \cite{Nac97}. 

Under these assumptions the following problems necessitating the evaluation
of two loops have been solved: The calculation of the gauge invariant color
field density \( \sum _{a}E_{k}^{a}(x)^{2} \), \( \sum _{a}B_{k}^{a}(x)^{2} \)
\cite{RD95} and of total or diffractive cross sections in soft high energy
reactions \cite{KD91}, \cite{Nac97}, \cite{dosch2}, \cite{BN99}. 

In the next section we investigate in which way the very consistent results
of the Gaussian model could be modified by the presence of higher cumulants.
It is not astonishing that in that case the program cannot be performed as far
as in the Gaussian approximation. We therefore concentrate on the leading effects
of some higher cumulants.

\section{Higher Cumulants}\label{highercumuls}

\subsection{Choice of the cumulants}

In order to relax the factorization hypothesis of the MSV which only takes into
account cumulants of order two, we introduce higher cumulants. The cumulant
expression for a correlator of fourth order is given by:
\begin{eqnarray}
\left\langle F_{\mu _{1}\nu _{1}}^{a_{1}}F_{\mu _{2}\nu _{2}}^{a_{2}}F_{\mu _{3}\nu _{3}}^{a_{3}}F_{\mu _{4}\nu _{4}}^{a_{4}}\right\rangle  & = & \left\langle \left\langle F_{\mu _{1}\nu _{1}}^{a_{1}}F_{\mu _{2}\nu _{2}}^{a_{2}}\right\rangle \right\rangle \left\langle \left\langle F_{\mu _{3}\nu _{3}}^{a_{3}}F_{\mu _{4}\nu _{4}}^{a_{4}}\right\rangle \right\rangle \nonumber \\
 & + & \left\langle \left\langle F_{\mu _{1}\nu _{1}}^{a_{1}}F_{\mu _{3}\nu _{3}}^{a_{3}}\right\rangle \right\rangle \left\langle \left\langle F_{\mu _{2}\nu _{2}}^{a_{2}}F_{\mu _{4}\nu _{4}}^{a_{4}}\right\rangle \right\rangle \nonumber \\
 & + & \left\langle \left\langle F_{\mu _{1}\nu _{1}}^{a_{1}}F_{\mu _{4}\nu _{4}}^{a_{4}}\right\rangle \right\rangle \left\langle \left\langle F_{\mu _{2}\nu _{2}}^{a_{2}}F_{\mu _{3}\nu _{3}}^{a_{3}}\right\rangle \right\rangle \nonumber \\
 & + & \left\langle \left\langle F_{\mu _{1}\nu _{1}}^{a_{1}}F_{\mu _{2}\nu _{2}}^{a_{2}}F_{\mu _{3}\nu _{3}}^{a_{3}}F_{\mu _{4}\nu _{4}}^{a_{4}}\right\rangle \right\rangle .\label{4korrentwicklung} 
\end{eqnarray}
We are not striving to find the most general ansatz for the fourth cumulant
since a full inclusion of all fourth cumulants compatible with the general requirements
of Lorentz and gauge covariance would introduce many new unknown parameters
into the model. Even for the local condensate there are six independent ones
\cite{bagan1}. Any quantitative computation would lose its predictive power.
Therefore we restrict ourselvers to studying the influence of some possible
cumulants in a qualitative way. Of course the cumulants are not independent
of each other, but there are some relations between them due to the equations
of motion \cite{bagan1}, \cite{Sim92}. However these relations are not sufficient
to determine the higher cumulants through the lower ones. 

Two non-abelian example structures for a fourth cumulant, which satisfy the
requirements of gauge and Lorentz invariance and which satisfy the normalisation
condition:
\begin{equation}
\left\langle \left\langle g^{4}F_{\mu _{1}\nu _{1}}^{a_{1}}F_{\mu _{1}\nu _{1}}^{a_{1}}F_{\mu _{2}\nu _{2}}^{a_{2}}F_{\mu _{2}\nu _{2}}^{a_{2}}\right\rangle \right\rangle =\mathcal{N}\left( \left\langle g^{2}FF\right\rangle \right) ^{2},
\end{equation}
 are: 
\begin{eqnarray}
\lefteqn {\left\langle \left\langle g^{4}F_{\mu _{1}\nu _{1}}^{a_{1}}(x_{1},w)F_{\mu _{2}\nu _{2}}^{a_{2}}(x_{2},w)F_{\mu _{3}\nu _{3}}^{a_{3}}(x_{3},w)F_{\mu _{4}\nu _{4}}^{a_{4}}(x_{4},w)\right\rangle \right\rangle =} & \hspace {2cm} & \nonumber \\
 &  & \frac{\mathcal{N}}{1152}\left( \left\langle g^{2}FF\right\rangle \right) ^{2}\nonumber\\
 &  & \Big \{f_{a_{1}a_{2}e}f_{a_{3}a_{4}e}(\epsilon _{\mu _{1}\nu _{1}\mu _{3}\nu _{3}}\epsilon _{\mu _{2}\nu _{2}\mu _{4}\nu _{4}}-\epsilon _{\mu _{1}\nu _{1}\mu _{4}\nu _{4}}\epsilon _{\mu _{2}\nu _{2}\mu _{3}\nu _{3}})\nonumber \\
 &  & +f_{a_{3}a_{1}e}f_{a_{2}a_{4}e}(\epsilon _{\mu _{1}\nu _{1}\mu _{4}\nu _{4}}\epsilon _{\mu _{2}\nu _{2}\mu _{3}\nu _{3}}-\epsilon _{\mu _{1}\nu _{1}\mu _{2}\nu _{2}}\epsilon _{\mu _{3}\nu _{3}\mu _{4}\nu _{4}})\nonumber \\
 &  & +f_{a_{2}a_{3}e}f_{a_{1}a_{4}e}(\epsilon _{\mu _{1}\nu _{1}\mu _{2}\nu _{2}}\epsilon _{\mu _{3}\nu _{3}\mu _{4}\nu _{4}}-\epsilon _{\mu _{1}\nu _{1}\mu _{3}\nu _{3}}\epsilon _{\mu _{2}\nu _{2}\mu _{4}\nu _{4}})\Big \}\nonumber \\
 &  & D(z_{1},z_{2},z_{3},z_{4},z_{5},z_{6})\label{4ansatz1} 
\end{eqnarray}
and
\begin{eqnarray}
\lefteqn {\left\langle \left\langle g^{4}F_{\mu _{1}\nu _{1}}^{a_{1}}(x_{1},w)F_{\mu _{2}\nu _{2}}^{a_{2}}(x_{2},w)F_{\mu _{3}\nu _{3}}^{a_{3}}(x_{3},w)F_{\mu _{4}\nu _{4}}^{a_{4}}(x_{4},w)\right\rangle \right\rangle =}\hspace {2cm} &  & \nonumber \\
 &  & \frac{-\mathcal{N}}{9216}\left( \left\langle g^{2}FF\right\rangle \right) ^{2}\nonumber \\
 &  & \Big \{\delta _{a_{1}a_{2}}\delta _{a_{3}a_{4}}(\delta _{\mu _{1}\mu _{2}}\delta _{\nu _{1}\nu _{2}}-\delta _{\mu _{1}\nu _{2}}\delta _{\mu _{2}\nu _{1}})(\delta _{\mu _{3}\mu _{4}}\delta _{\nu _{3}\nu _{4}}-\delta _{\mu _{3}\nu _{4}}\delta _{\mu _{4}\nu _{3}})\nonumber \\
 &  & +\delta _{a_{1}a_{3}}\delta _{a_{2}a_{4}}(\delta _{\mu _{1}\mu _{3}}\delta _{\nu _{1}\nu _{3}}-\delta _{\mu _{1}\nu _{3}}\delta _{\mu _{3}\nu _{1}})(\delta _{\mu _{2}\mu _{4}}\delta _{\nu _{2}\nu _{4}}-\delta _{\mu _{2}\nu _{4}}\delta _{\mu _{4}\nu _{2}})\nonumber \\
 &  & +\delta _{a_{1}a_{4}}\delta _{a_{2}a_{3}}(\delta _{\mu _{1}\mu _{4}}\delta _{\nu _{1}\nu _{4}}-\delta _{\mu _{1}\nu _{4}}\delta _{\mu _{4}\nu _{1}})(\delta _{\mu _{2}\mu _{3}}\delta _{\nu _{2}\nu _{3}}-\delta _{\mu _{2}\nu _{3}}\delta _{\mu _{3}\nu _{2}})\Big \}\nonumber \\
 &  & D(z_{1},z_{2},z_{3},z_{4},z_{5},z_{6}).\label{4ansatz2} 
\end{eqnarray}
Here \( z_{i} \) are the differences between the coordinates \( z_{1}=x_{1}-x_{2},\, z_{2}=x_{1}-x_{3},\ldots ,\, z_{6}=x_{3}-x_{4} \).
The scalar function \( D \) falls off fast if any of the distances \( |z_{i}| \)
becomes large, it is normalized \( D(0,0,\ldots ,0)=1 \). We have extracted
the dimensionful quantity \( \left\langle g^{2}FF\right\rangle ^{2} \), the
number which determines the strength of the full correlator (\ref{4korrentwicklung}).
The first structure is similar to the four gluon vertex, \( f_{a_{1}a_{2}e} \)
being the structure constants of SU(3). The correlation lengths \( a_{i} \)
are defined by the following conditions:

\begin{equation}
\label{korrfunc}
\int _{0}^{\infty }D(0,\ldots ,z_{i},\ldots ,0)\, dz_{i}=a_{i}\quad i=1\ldots 6.
\end{equation}
 In the following we put \( a_{i}=a \) and all distances are taken to be in
units of the correlation length \( a \). In a Euclidean space we make a simple
Gaussian ansatz for \( D \):

\begin{equation}
\label{4korrfunc}
D(z_{1},z_{2},z_{3},z_{4},z_{5},z_{6})=e^{-(z_{1}^{2}+z^{2}_{2}+z_{3}^{2}+z_{4}^{2}+z_{5}^{2}+z_{6}^{2})/\lambda ^{2}},
\end{equation}
where \( \lambda =\frac{2}{\sqrt{\pi }}a \). As the distances \( z_{4} \),
\( z_{5} \) and \( z_{6} \) can be expressed as a function of \( z_{1} \),
\( z_{2} \) and \( z_{3} \), \( D \) is a function of the latter three independent
variables only. Ansatz (\ref{4korrfunc}) is not supposed to be a realistic
choice but rather a simple manageable expression taking into account the cluster
property.

\subsection{Single Wegner-Wilson loop: Potential}

In order to obtain the static potential of a quark-antiquark pair, the expectation
value of the Wegner-Wilson loop is calculated through the cumulant expansion
\cite{Dos87}, \cite{DS88}:
\begin{eqnarray}
\lefteqn {\left< \mathrm{Tr}\mathbf{W}[S]\right> =\mathrm{TrP}_{S}} &  & \nonumber \\
 &  & \exp \Big [-\frac{g^{2}}{4\cdot 2!}\int _{S}\int _{S}d\sigma _{\mu _{1}\nu _{1}}(x_{1})d\sigma _{\mu _{2}\nu _{2}}(x_{2})\, \left< \left< \, \mathbf{F}_{\mu _{1}\nu _{1}}(x_{1},w)\mathbf{F}_{\mu _{2}\nu _{2}}(x_{2},w)\, \right> \right> \nonumber \\
 &  & +\frac{g^{4}}{2^{4}\cdot 4!}\int _{S}\int _{S}\int _{S}\int _{S}d\sigma _{\mu _{1}\nu _{1}}(x_{1})d\sigma _{\mu _{2}\nu _{2}}(x_{2})d\sigma _{\mu _{3}\nu _{3}}(x_{3})d\sigma _{\mu _{4}\nu _{4}}(x_{4})\nonumber \\
 &  & \hspace {3cm}\cdot \left< \left< \, \mathbf{F}_{\mu _{1}\nu _{1}}(x_{1},w)\mathbf{F}_{\mu _{2}\nu _{2}}(x_{2},w)\mathbf{F}_{\mu _{3}\nu _{3}}(x_{3},w)\mathbf{F}_{\mu _{4}\nu _{4}}(x_{4},w)\, \right> \right> \nonumber \\
 &  & -\frac{g^{6}}{2^{6}\cdot 6!}\ldots \Big ].\label{wwloopexpanded} 
\end{eqnarray}
 where \( S \) is the integration surface enclosed by the loop and \( \mathrm{P}_{S} \)
denotes surface ordering. 

The first ansatz (\ref{4ansatz1}) gives the following contribution to the potential,
after performing the trace and summing over all color indices: 
\begin{eqnarray}
V^{(1)}_{4} & = & \lim _{T\rightarrow \infty }\frac{1}{T}\mathrm{P}_{S}\nonumber \\
 &  & \frac{1}{2^{4}\cdot 4!}\int _{S}\int _{S}\int _{S}\int _{S}d\sigma _{\mu _{1}\nu _{1}}(x_{1})d\sigma _{\mu _{2}\nu _{2}}(x_{2})d\sigma _{\mu _{3}\nu _{3}}(x_{3})d\sigma _{\mu _{4}\nu _{4}}(x_{4})\nonumber \\
 &  & \cdot \frac{1}{128}(-\epsilon _{\mu _{1}\nu _{1}\mu _{4}\nu _{4}}\epsilon _{\mu _{2}\nu _{2}\mu _{3}\nu _{3}}-\epsilon _{\mu _{1}\nu _{1}\mu _{2}\nu _{2}}\epsilon _{\mu _{3}\nu _{3}\mu _{4}\nu _{4}}+2\epsilon _{\mu _{1}\nu _{1}\mu _{3}\nu _{3}}\epsilon _{\mu _{2}\nu _{2}\mu _{4}\nu _{4}})\nonumber \\
 &  & \mathcal{N}\left( \left\langle g^{2}FF\right\rangle \right) ^{2}D(z_{1},z_{2},z_{3},z_{4},z_{5},z_{6}).
\end{eqnarray}
Asuming that the Wegner-Wilson loop lies in a plane with the reference point
also on this plane, the integration surface \( S \) is just the rectangle enclosed
by the loop. It is easy to see that the \( \epsilon  \)-tensors vanish on this
plane and that the cumulant of the first ansatz (\ref{4ansatz1}) does not contribute
to the potential of the \( q\bar{q} \)-pair. 

\begin{figure}[h]
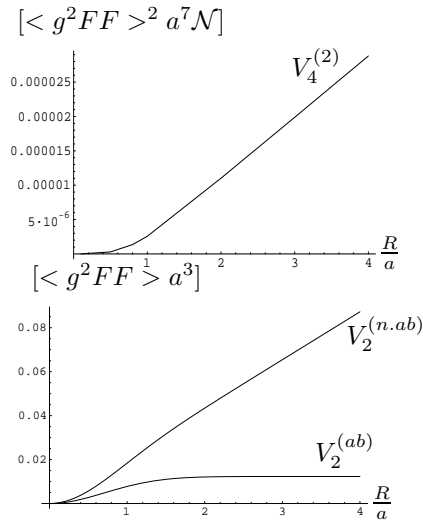

\centering\begin{minipage}{12cm} 
\begin{minipage}{6cm}
\centering
\input{potential4normed.pstex_t}
\end{minipage}
\begin{minipage}{6cm}\centering
\input{potential2normed.pstex_t}
\end{minipage}

\caption[Fourth cumulant contribution to the potential]{\label{figpot4}On the
left the numerical result for the fourth cumulant contribution (\protect\ref{4ansatz2})
to the potential \( V^{(2)}_{4} \) of a static \( q\bar{q} \)-pair is shown.
For large quark separations \( R \) a linear rise is obtained whereas for small
separation a \( R^{4} \) behaviour is found. On the right the results of the
Gaussian model for the Abelian (ab) and non-Abelian (n.ab) case are shown.}

\end{minipage}
\end{figure}

Inserting ansatz (\ref{4ansatz2}) and summing over all color indices gives
the following fourth cumulant contribution to the potential:
\begin{eqnarray}
V^{(2)}_{4} & = & \lim _{T\rightarrow \infty }\frac{1}{T}\mathrm{P}_{S}\nonumber \\
 &  & \frac{1}{9216\cdot 2^{4}\cdot 4!}\int _{S}\int _{S}\int _{S}\int _{S}d\sigma _{\mu _{1}\nu _{1}}(x_{1})d\sigma _{\mu _{2}\nu _{2}}(x_{2})d\sigma _{\mu _{3}\nu _{3}}(x_{3})d\sigma _{\mu _{4}\nu _{4}}(x_{4})\nonumber \\
 &  & \Big \{16(\delta _{\mu _{1}\mu _{2}}\delta _{\nu _{1}\nu _{2}}-\delta _{\mu _{1}\nu _{2}}\delta _{\mu _{2}\nu _{1}})(\delta _{\mu _{3}\mu _{4}}\delta _{\nu _{3}\nu _{4}}-\delta _{\mu _{3}\nu _{4}}\delta _{\mu _{4}\nu _{3}})\nonumber \\
 &  & -2(\delta _{\mu _{1}\mu _{3}}\delta _{\nu _{1}\nu _{3}}-\delta _{\mu _{1}\nu _{3}}\delta _{\mu _{3}\nu _{1}})(\delta _{\mu _{2}\mu _{4}}\delta _{\nu _{2}\nu _{4}}-\delta _{\mu _{2}\nu _{4}}\delta _{\mu _{4}\nu _{2}})\nonumber \\
 &  & +16(\delta _{\mu _{1}\mu _{4}}\delta _{\nu _{1}\nu _{4}}-\delta _{\mu _{1}\nu _{4}}\delta _{\mu _{4}\nu _{1}})(\delta _{\mu _{2}\mu _{3}}\delta _{\nu _{2}\nu _{3}}-\delta _{\mu _{2}\nu _{3}}\delta _{\mu _{3}\nu _{2}})\nonumber \\
 &  & \mathcal{N}\left( \left\langle g^{2}FF\right\rangle \right) ^{2}D(z_{1},z_{2},z_{3},z_{4},z_{5},z_{6})\Big \}.
\end{eqnarray}
After suitable parametrisation of the rectangular integration surface bounded
by the loop, a non-vanishing contribution to the potential is obtained \cite{kornelis1}:
\begin{eqnarray}
V^{(2)}_{4} & = & \lim _{T\rightarrow \infty }\frac{5}{3\cdot 2^{25}N_{C}}\mathcal{N}\left( \left\langle g^{2}FF\right\rangle \right) ^{2}\nonumber \\
 &  & \int _{-1}^{1}d\tau _{1}\int _{-1}^{1}d\tau _{2}\int _{-1}^{1}d\tau _{3}\int _{-1}^{1}d\tau _{4}\int _{-1}^{1}ds_{1}\int _{-1}^{1}ds_{2}\int _{-1}^{1}ds_{3}\int _{-1}^{1}ds_{4}\nonumber \\
 &  & \cdot T^{3}R^{4}\exp \bigg [-R^{2}\Big [(s_{2}-s_{1})^{2}+(s_{3}-s_{2})^{2}+(s_{4}-s_{3})^{2}\nonumber \\
 &  & \hspace {2.8cm}+(s_{4}-s_{2})^{2}+(s_{3}-s_{1})^{2}+(s_{4}-s_{1})^{2}\Big ]\nonumber \\
 &  & \hspace {1.9cm}-T^{2}\Big [(\tau _{2}-\tau _{1})^{2}+(\tau _{3}-\tau _{2})^{2}+(\tau _{4}-\tau _{3})^{2}\nonumber \\
 &  & \hspace {2.8cm}+(\tau _{4}-\tau _{2})^{2}+(\tau _{3}-\tau _{1})^{2}+(\tau _{4}-\tau _{1})^{2}\Big ]\bigg ].
\end{eqnarray}
The integrations and the limit \( T\rightarrow \infty  \) can be performed
numerically and the result is shown in figure \ref{figpot4}. The fourth cumulant
contribution to the potential is plotted as a function of the quark-antiquark
distance \( R \). For large \( R \) a linear rise is found, as in the Gaussian
model. For short distances the potential behaves as \( R^{4} \), in contrast
to the Gaussian model, where the short distance behaviour is like \( R^{2} \). 

\subsection{The field of a static \protect\( q\bar{q}\protect \)-pair}

We consider a static quark-antiquark pair in a color singlet state at fixed
space points. It gives rise to a chromoelectric and chromomagnetic field whose
components form the field strength tensor. In order to calculate the squared
field strength we introduce a plaquette \( \mathbf{P}_{\mu \nu }(x) \), representing
a small Wegner-Wilson loop in the \( \mu \nu  \)-plane with center point \( x \)
and dimension \( R_{P} \) (fig. \ref{plakette}). In the limit \( Rp\rightarrow 0 \)
its expansion gives the following relationship with the gauge invariant field
density \cite{RD95}:
\begin{eqnarray}
\mathbf{P}_{\mu \nu }(x)=N_{C}-\frac{1}{4}R_{P}^{4}g^{2}\sum _{a}F^{a}_{\mu \nu }F^{a}_{\mu \nu }+\mathcal{O}(R_{P}^{8}) & 
\end{eqnarray}
where there is no summation over \( \mu  \) and \( \nu  \). This plaquette
is evaluated in the presence of a static \( q\bar{q} \)-pair whose world lines
and color connectors form in turn a large rectangular loop of width \( R \)
and length \( T \) going to infinity. We thus consider the quantity:
\begin{equation}
\label{gluonfeld}
f_{\mu \nu }(x):=\frac{4}{R_{P}^{4}g^{2}}\frac{\left< \mathrm{Tr}\mathbf{W}[\partial S]\mathrm{Tr}\mathbf{P}_{\mu \nu }(x)\right> -\left< \mathrm{Tr}\mathbf{W}[\partial S]\right> \left< \mathrm{Tr}\mathbf{P}_{\mu \nu }(x)\right> }{\left< \mathrm{Tr}\mathbf{W}[\partial S]\right> }
\end{equation}
 and
\begin{equation}
\label{ftensor}
\lim _{R_{P}\rightarrow 0}f_{\mu \nu }(x)=\left( \begin{array}{cccc}
0 & -B_{z}^{2} & -B_{y}^{2} & -E_{x}^{2}\\
-B_{z}^{2} & 0 & -B_{x}^{2} & -E_{y}^{2}\\
-B_{y}^{2} & -B_{x}^{2} & 0 & -E_{z}^{2}\\
-E_{x}^{2} & -E_{y}^{2} & -E_{z}^{2} & 0
\end{array}\right) ,
\end{equation}
 \( \vec{E} \) and \( \vec{B} \) being the chromoelectric and chromomagnetic
field. The coordinates of the plaquette are \( x \), \( y \) and \( z \):
because of rotational symmetry in the \( x_{1} \)-\( x_{2} \)-plane, we choose
\( y=0 \). Assuming that the Wegner-Wilson loop lies in the \( x_{3} \)-\( x_{4} \)-plane,
\( x \) is the perpendicular distance to the quark axis, whereas \( z \) represents
the distance from the origin parallel to the quark axis see fig. \ref{plakette}. 
\begin{figure}
\centering\begin{minipage}{10cm}\centering\input{plakette.pstex_t}

\caption[Wegner-Wilson loop and plaquette $P_{\mu\nu}$]{\label{plakette}Configuration
of the Wegner-Wilson loop \( W \) and the plaquette \( P \). The quark-antiquark
axis is oriented in the \( x_{3} \)-direction. The plaquette distance from
the origin parallel to the quark-antiquark axis is represented by \( z \) and
the distance perpendicular to the quark axis is \( x \). In this figure the
plaquette \( P_{34} \) is shown.}

\end{minipage}
\end{figure}
 We proceed by studying \( f_{\mu \nu } \) in the limit \( R_{P}\rightarrow 0 \).
After the expansion of the exponentials in (\ref{gluonfeld}) we obtain:
\begin{eqnarray}
\lefteqn {f(x)=\frac{4}{R_{P}^{4}g^{2}}\frac{1}{\left\langle \mathrm{Tr}\mathbf{W}\right\rangle }} &  & \nonumber \\
 &  & \cdot \Bigg (\sum _{n=1}^{\infty }\frac{(-i)^{n}}{2^{n}n!}\, \stackrel{\mbox {ordered}}{\int \ldots \int }d\sigma ^{W}_{\mu _{1}\nu _{1}}\ldots d\sigma ^{W}_{\mu _{n}\nu _{n}}\, \mathrm{Tr}\left[ \mathbf{T}^{a_{1}}\ldots \mathbf{T}^{a_{n}}\right] \stackrel{\mbox {free}}{\int \int }d\sigma _{\mu \nu }^{P}d\sigma _{\rho \sigma }^{P}\nonumber \\
 &  & \hspace {2cm}\cdot \frac{(-i)^{2}}{2^{2}2!}\mathrm{Tr}\left[ \mathbf{T}^{a}\mathbf{T}^{b}\right] \left\langle g^{n}F^{a_{1}}_{\mu _{1}\nu _{1}}\cdots F^{a_{n}}_{\mu _{n}\nu _{n}}g^{2}F^{a}_{\mu \nu }F^{b}_{\rho \sigma }\right\rangle -\nonumber \\
 &  & \hfill -\sum _{n=1}^{\infty }\frac{(-i)^{n}}{2^{n}n!}\, \stackrel{\mbox {ordered}}{\int \ldots \int }d\sigma ^{W}_{\mu _{1}\nu _{1}}\ldots d\sigma ^{W}_{\mu _{n}\nu _{n}}\, \mathrm{Tr}\left[ \mathbf{T}^{a_{1}}\ldots \mathbf{T}^{a_{n}}\right] \stackrel{\mbox {free}}{\int \int }d\sigma _{\mu \nu }^{P}d\sigma _{\rho \sigma }^{P}\nonumber \\
 &  & \hspace {2cm}\cdot \frac{(-i)^{2}}{2^{2}2!}\mathrm{Tr}\left[ \mathbf{T}^{a}\mathbf{T}^{b}\right] \left\langle g^{n}F^{a_{1}}_{\mu _{1}\nu _{1}}\cdots F^{a_{n}}_{\mu _{n}\nu _{n}}\right\rangle \left\langle g^{2}F^{a}_{\mu \nu }F^{b}_{\rho \sigma }\right\rangle \Bigg ),\label{4gluonfeld} 
\end{eqnarray}
where the integrations over the surface bounded by the Wegner-Wilson loop are
surface ordered. The matrix elements of \( f \) are given by the different
orientations of the plaquette. In the Gaussian model it was possible to insert
the factorizations and rewrite the summations in such a way as to perform the
summations \cite{RD95}. Here this full summation is not feasible. Nevertheless
to get an idea about what kind of contribution a fourth cumulant could give
to the fields, we confine ourselves to the lowest non-trivial term. It should
give a qualitative indication of the effects of higher cumulants. After inserting
the cumulant definition (\ref{4korrentwicklung}) we obtain the leading contribution
from the fourth cumulant:
\begin{eqnarray}
\lefteqn {\delta f:=\frac{4}{R_{P}^{4}g^{2}}} &  & \nonumber \\
 &  & \cdot \frac{1}{2^{2}2!}\stackrel{\mbox {ordered}}{\int \int }d\sigma ^{W}_{\mu _{1}\nu _{1}}(x_{W})d\sigma ^{W}_{\mu _{2}\nu _{2}}(x_{W}')\mathrm{Tr}\left[ \mathbf{T}^{a_{1}}\mathbf{T}^{a_{2}}\right] \stackrel{\mbox {free}}{\int \int }d\sigma _{\mu \nu }^{P}(x_{P})d\sigma _{\rho \sigma }^{P}(x_{P}')\nonumber \\
 &  & \cdot \frac{1}{2^{2}2!}\mathrm{Tr}\left[ \mathbf{T}^{a}\mathbf{T}^{b}\right] \left\langle \left\langle g^{2}F^{a_{1}}_{\mu _{1}\nu _{1}}(x_{W})F^{a_{2}}_{\mu _{2}\nu _{2}}(x_{W}')g^{2}F^{a}_{\mu \nu }(x_{P})F^{b}_{\rho \sigma }(x_{P}')\right\rangle \right\rangle .
\end{eqnarray}
Since we are mainly interested in the \( x,z \)-dependence of \( \delta f \)
we omit the \( x \) and \( z \)-independent normalization \( \left\langle \mathrm{Tr}\mathbf{W}\right\rangle  \). 

To compute the contribution of each ansatz (\ref{4ansatz1}) or (\ref{4ansatz2})
the traces must be performed and a summation over all color indices must be
carried out. After suitable parametrization of the plane surfaces bounded by
the Wegner-Wilson loop and the plaquette, we obtain for the contribution of
the first ansatz (\ref{4ansatz1}) \cite{kornelis1}:
\begin{eqnarray}
\delta f_{1} & = & \frac{-N_{C}}{18432g^{2}}\mathcal{N}\left( \left\langle g^{2}FF\right\rangle \right) ^{2}\left( \begin{array}{cccc}
0 & 1 & 0 & 0\\
1 & 0 & 0 & 0\\
0 & 0 & 0 & 0\\
0 & 0 & 0 & 0
\end{array}\right) \xi (x,z,R),
\end{eqnarray}
and for the second ansatz (\ref{4ansatz2}) \cite{kornelis1}:
\begin{eqnarray}
\delta f_{2} & = & \frac{-1}{147456g^{2}}\mathcal{N}\left( \left\langle g^{2}FF\right\rangle \right) ^{2}\left( \begin{array}{cccc}
0 & 4 & 4 & 4\\
4 & 0 & 4 & 4\\
4 & 4 & 0 & 5\\
4 & 4 & 5 & 0
\end{array}\right) \xi (x,z,R).
\end{eqnarray}
where both contributions have the same \( x,z \)-dependence: 
\begin{eqnarray}
\lefteqn {\xi (x,z,R):=\int _{-1}^{1}R^{2}ds_{W}ds_{W}'\int ^{\infty }_{-\infty }d\tau _{W}d\tau _{W}'} &  & \nonumber \\
 &  & \cdot \exp \bigg [-\frac{1}{\lambda ^{2}}\Big [4x^{2}+2(z-\frac{1}{2}Rs_{W})^{2}+2(z-\frac{1}{2}Rs_{W}')^{2}\nonumber \\
 &  & \hspace {2cm}+\frac{1}{4}R^{2}(s_{W}'-s_{W})^{2}+2(\tau ^{2}_{W}+\tau _{W}'^{2})+(\tau _{W}'-\tau _{W})^{2}\Big ]\bigg ].
\end{eqnarray}
The first ansatz (\ref{4ansatz1}) only contributes to the chromomagnetic field
in the direction parallel to the quark axis:
\begin{eqnarray}
\delta B_{z}^{2} & = & \frac{N_{C}}{18432g^{2}}\mathcal{N}\left( \left\langle g^{2}FF\right\rangle \right) ^{2}\xi (x,z,R),\label{fieldres1} 
\end{eqnarray}
whereas the second ansatz (\ref{4ansatz2}) contributes to all components of
both the chromoelectric and the chromomagnetic field:
\begin{eqnarray}
\delta B_{x}^{2}=\delta B_{y}^{2}=\delta B_{z}^{2} & = & \frac{4}{147456g^{2}}\mathcal{N}\left( \left\langle g^{2}FF\right\rangle \right) ^{2}\xi (x,z,R)\nonumber \\
\delta E_{x}^{2}=\delta E_{y}^{2} & = & \frac{4}{147456g^{2}}\mathcal{N}\left( \left\langle g^{2}FF\right\rangle \right) ^{2}\xi (x,z,R)\nonumber \\
\delta E_{z}^{2} & = & \frac{5}{147456g^{2}}\mathcal{N}\left( \left\langle g^{2}FF\right\rangle \right) ^{2}\xi (x,z,R).\label{fieldres2} 
\end{eqnarray}

\begin{figure}
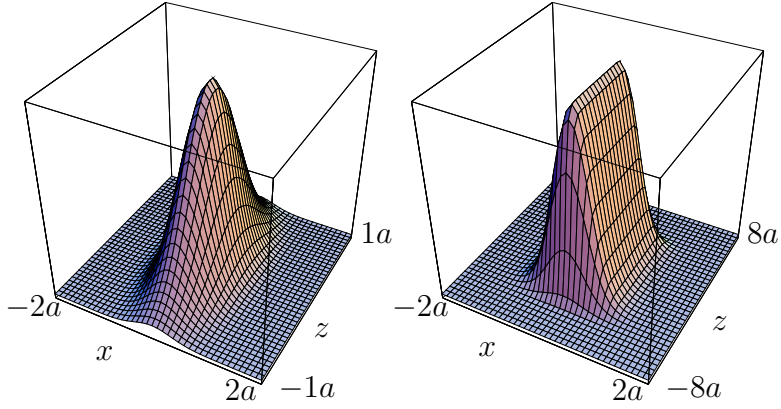

\centering
\begin{minipage}{10cm}
\centering\begin{tabular}{cc}
\begin{minipage}{5cm}
\centering\input{schlauch_r1.pstex_t} 
\end{minipage}&
 \begin{minipage}{5cm} 
\centering\input{schlauch_r8.pstex_t} 
\end{minipage} 
\end{tabular}
 \caption[The function $\xi(x,z,R)$ for $R=1,8$]{\label{schlauchfig}The function
\( \xi (x,z,R) \) in arbitrary units for the \( q\bar{q} \)-separations \( R=1 \)
and \( R=8 \). The quarks have coordinates \( x=0 \) and \( z=\pm 0.5a \)
for \( R=1 \) and \( x=0 \) and \( z=\pm 4a \) for \( R=8 \). One observes
the formation of a flux tube, note the different \( z \)-axis scales of the
two plots.} 
\end{minipage}
 \end{figure}

The numerical evaluation of the \( x,z \)-dependence \( \xi (x,z,R) \) of
the fourth cumulant contributions to the fields is shown in figure \ref{schlauchfig}
for the \( q\bar{q} \)-separations \( R=1 \) and \( R=8 \).
The result is that, for increasing separation between the \( q\bar{q} \)-pair,
a flux tube is formed. With (\ref{fieldres1}) and (\ref{fieldres2}) the following
conclusions can be drawn:

\begin{itemize}
\item The first ansatz (\ref{4ansatz1}) leads to the formation of a purely chromomagnetic
flux tube for the \( z \)-component of the field, but to no area law of the
Wegner-Wilson loop. 
\item The contribution (\ref{fieldres2}) of the second ansatz (\ref{4ansatz2}) can
be split into a deformation of the vacuum field density with no preferred field
components and a flux tube for the color-electric field in the \( z \)-direction.
Of course both contribute to the energy density of the system. In this case
confinement also reflects itself in the area law. 
\end{itemize}
We thus find for the first ansatz (\ref{4ansatz1}) a contribution to a chromomagnetic
flux tube which was not present in the Gaussian approximation. The easiest interpretation
of this behaviour is that the flux tube has been created by a pair of magnetic
monopoles. At first sight there seems to be a contradiction between the formation
of a flux tube and the absence of a contribution to the area law of the Wegner-Wilson
loop. However in the Wegner-Wilson loop the exponent \( g\mathbf{A}_{\mu }dx^{\mu } \)
only takes into account the interaction of a color-\emph{electric} charge. Therefore
it is not astonishing that the area law of the Wegner-Wilson loop (\ref{wwloop})
is not a criterium for confinement of color-\emph{magnetic} monopoles.

\subsection{High Energy Scattering}

\begin{figure}[h]
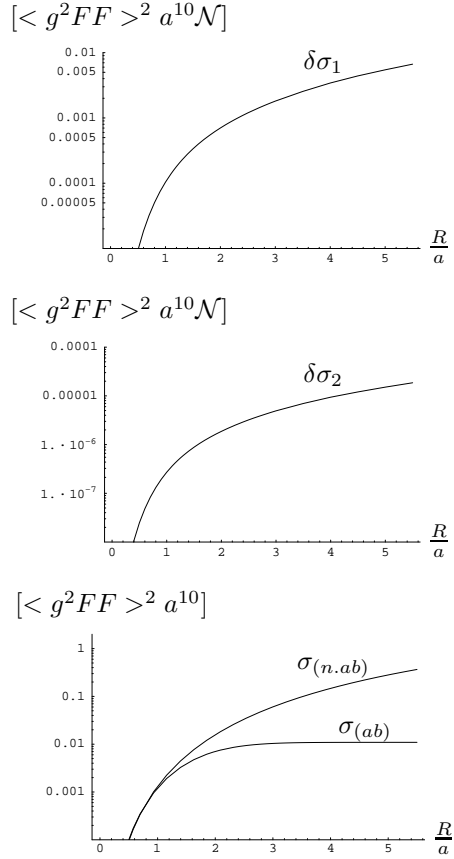

\centering\begin{minipage}{12cm} 
\vspace{0.5cm}
\begin{minipage}{6cm}
\centering
\input{fig_eq_log1.pstex_t}
\vspace{0.5cm}
\end{minipage}
\begin{minipage}{6cm}\centering
\input{fig_eq_log2.pstex_t}
\vspace{0.5cm} 
\end{minipage}
 \begin{minipage}{6cm}
\centering
\input{fig_eq_log_gaussian.pstex_t}
\end{minipage} 
\caption[Fourth cumulant contributions to the total cross section]{\label{e_overview4}
The first two figures give the fourth cumulant contributions to the total dipole-dipole
scattering cross section for ansatz (\protect\ref{4ansatz1}) and (\protect\ref{4ansatz2})
respectively. The last figure gives the results of the Gaussian model for the
non-Abelian (n.ab) and the Abelian (ab) case.}

\end{minipage}
\end{figure}

Soft hadron-hadron high energy scattering has been extensively studied using
the Gaussian model \cite{Dos94}, \cite{Sim96}, \cite{Dos96}, \cite{Nac97},
\cite{BN99}. We have investigated the effect of relaxing the Gaussian approximation
in much the same way as for the static \( q\bar{q} \)-pair. Both fourth cumulants
(\ref{4ansatz1}) and (\ref{4ansatz2}) can be continued to Minkowski space
in exactly the same way as the second cumulant and in the final results only
the separations in transverse space enter; this means that the Euclidean expressions
for the correlators can be used. In fig. \ref{e_overview4} we give the total
dipole-dipole cross section as a function of the dipole size for two equal dipoles.
The qualitative behaviour is the same as for the Gaussian model: at short distances
the cross section is proportional to the fourth power of the size of the dipoles,
for large sizes it approaches a quadratic dependence. We have also studied the
dependence of the cross section on dipole size with one fixed-size dipole. Again
we find the same behaviour as in the Gaussian model namely for small sizes a
dependence on the square of dipole size and for large sizes an approach to a
linear behaviour. 

\section{Summary and outlook}

We have found that the contributions of higher cumulants in the cluster expansion
do not lead to a change of the qualitative behaviour of phenomenological results
of the Gaussian model. There are however interesting new features in the field
configurations of a static \( q\bar{q} \)-pair. The resulting flux tube of
ansatz (\ref{4ansatz2}) consists of an increased isotropic vacuum density inside
the region described by the function \( \xi (x,z,R) \) and in addition of a
color-electric flux tube. The ansatz (\ref{4ansatz1}) leads to the formation
of a color-magnetic flux tube, very similar to the color-electric flux tube
of the Gaussian model. The easiest interpretation of such a flux tube is that
it has been generated by a pair of confined monopoles. Though a static bare
quark carries only color-electric charge it is not impossible that it acquires
a magnetic charge in the vacuum. For a color-magnetic monopole without color-electric
charge the area law of the Wegner-Wilson loop as evaluated in the MSV is then
no longer an indication of confinement. It would therefore be very interesting
to investigate higher cumulants on the lattice since they might reveal new features
of non-Abelian gauge theories.

\section*{Acknowledgements}

The authors want to thank E. Berger, N. Brambilla, D. Gromes, O. Nachtmann,
J. Polonyi and A. Vairo for interesting discussions. 

\bibliographystyle{revtex}
\bibliography{myrefs,lithd,bibnew}

\end{document}